\documentclass[aps,pra,reprint,showpacs,floatfix,longbibliography]{revtex4-1}
\usepackage{graphicx, amsmath, amssymb, hyperref, multirow}
\usepackage[caption=false]{subfig}

\allowdisplaybreaks[1] 

\begin{document}


\newcommand{\braket}[2]{{\left\langle #1 \middle| #2 \right\rangle}}
\newcommand{\bra}[1]{{\left\langle #1 \right|}}
\newcommand{\ket}[1]{{\left| #1 \right\rangle}}
\newcommand{\ketbra}[2]{{\left| #1 \middle\rangle \middle \langle #2 \right|}}


\title{Quantum Walk Search on the Complete Bipartite Graph}

\author{Mason L.~Rhodes}
	\email{masonrhodes@creighton.edu}
\author{Thomas G.~Wong}
	\email{thomaswong@creighton.edu}
	\affiliation{Department of Physics, Creighton University, 2500 California Plaza, Omaha, NE 68178}

\begin{abstract}
	The coined quantum walk is a discretization of the Dirac equation of relativistic quantum mechanics, and it is the basis of many quantum algorithms. We investigate how it searches the complete bipartite graph of $N$ vertices for one of $k$ marked vertices with different initial states. We prove intriguing dependence on the number of marked and unmarked vertices in each partite set. For example, when the graph is irregular and the initial state is the typical uniform superposition over the vertices, then the success probability can vary greatly from one timestep to the next, even alternating between 0 and 1, so the precise time at which measurement occurs is crucial. When the initial state is a uniform superposition over the edges, however, the success probability evolves smoothly. As another example, if the complete bipartite graph is regular, then the two initial states are equivalent. Then if two marked vertices are in the same partite set, the success probability reaches $1/2$, but if they are in different partite sets, it instead reaches $1$. This differs from the complete graph, which is the quantum walk formulation of Grover's algorithm, where the success probability with two marked vertices is $8/9$. This reveals a contrast to the continuous-time quantum walk, whose evolution is governed by Schr\"odinger's equation, which asymptotically searches the regular complete bipartite graph with any arrangement of marked vertices in the same manner as the complete graph.
\end{abstract}


\maketitle


\section{Introduction}

The coined quantum walk is a spatial and temporal discretization of the Dirac equation of relativistic quantum mechanics \cite{Meyer1996a,Meyer1996b,Aharonov2001}, and it is a fundamental method for designing quantum algorithms \cite{Ambainis2003}, such as for searching \cite{SKW2003}, solving element distinctness \cite{Ambainis2004}, and evaluating boolean formulas \cite{Ambainis2010}. They are universal for quantum computing \cite{Lovett2010}, meaning any quantum algorithm can be formulated as a quantum walk, and they have been implemented in a variety of physical systems, including nuclear magnetic resonance \cite{Ryan2005}, optical waveguides \cite{Perets2008}, trapped atoms \cite{Karski2009}, and trapped ions \cite{Schmitz2009}.

In a coined quantum walk, the $N$ vertices of a graph label orthonormal basis states $\{ \ket{1}, \ket{2}, \dots, \ket{N} \}$ of an $N$-dimensional vertex Hilbert space. The walker also has an internal coin or spin degree of freedom indicating which direction the particle is pointing. Together, $\ket{u} \otimes \ket{v}$ or $\ket{uv}$ denotes a particle at vertex $u$ pointing to vertex $v$. A step of the quantum walk is $U_\text{walk} = SC$. Here, $C$ is the Grover diffusion coin \cite{SKW2003} that inverts the amplitudes of the coin states of each vertex about their average, and $S$ is the flip-flop shift \cite{AKR2005} that causes the particle to hop and turn around, so $S \ket{uv} = \ket{vu}$. For a detailed definition of the coined quantum walk for both regular and irregular graphs, see \cite{Wong27}.

Quantum walks are often used to explore how quantum computers, or quantum cellular automata, search a graph or network for marked nodes by querying an oracle $Q$ \cite{SKW2003,AKR2005,CG2004}, which flips the sign of the amplitudes at marked vertices. So the system evolves by repeated applications of
\begin{equation}
	\label{eq:U}
	U = U_\text{walk} Q = SCQ.
\end{equation}
Typically, the initial state is chosen to be a uniform superposition over the vertices, so if one measures the initial state, the walker is found at each vertex with equal probability. If the graph is complete, meaning each vertex is adjacent to every other, the graph is equivalent to the unordered database of Grover's algorithm \cite{Grover1996}. As in Grover's algorithm, search using a quantum walk is accomplished in $O(\sqrt{N})$ time, or if there are $k$ marked vertices, in $O(\sqrt{N/k})$ time \cite{AKR2005,CG2004,Wong10}.

\begin{figure}
\begin{center}
	\includegraphics{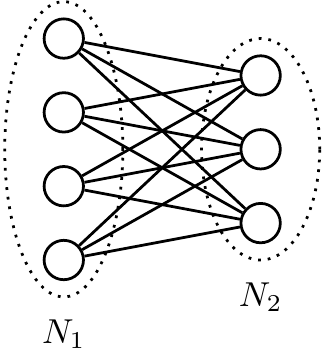}
	\caption{\label{fig:bipartite_unmarked} A complete bipartite graph with $N_1 = 4$ and $N_2 = 3$ vertices in each partite set.}
\end{center}
\end{figure}

In this paper, we consider search on the complete bipartite graph, an example of which is shown in Fig.~\ref{fig:bipartite_unmarked}. A bipartite graph can be partitioned into two vertex sets $X$ and $Y$ with $N_1$ and $N_2$ vertices, respectively, such that vertices in the same vertex set are not adjacent to each other. If the bipartite graph is complete, then each vertex in one partite set is adjacent to all vertices in the other vertex set.

Search on the complete bipartite graph using a continuous-time quantum walk, where the evolution is governed by Schr\"odinger's equation, has been previously explored in various cases. If the graph is regular, meaning $N_1 = N_2 = N/2$, then it is a strongly regular graph, and search on it with a single marked vertex was considered in \cite{Wong5}. Since it is regular, the graph Laplacian and adjacency matrix effect the same walk. If the graph is irregular, meaning $N_1 \ne N_2$, then search on it with the adjacency matrix and a single marked vertex was considered in \cite{Novo2015}. In \cite{Wong19}, search with multiple marked vertices and both the Laplacian and adjacency matrix were investigated, and they showed that when walking with the adjacency matrix, a non-uniform initial state is superior to the typical uniform one. Finally, search on the complete bipartite graph was also explored using a scattering quantum walk \cite{Reitzner2009}, and while their choice of parameters makes it equivalent to the coined quantum walk considered here, their initial state places the particle uniformly across partite set $X$, pointing to set $Y$, so it does not start in the $Y$ vertices at all.

Here, we consider two initial states. The first is the typical uniform superposition over all vertices $\ket{s}$. Then, if the position of the walker is measured, it has an equal probability of $1/N$ of being found at each vertex. This expresses our initial lack of information as to which vertex is marked, so they are guessed with equal probability. Since the discrete-time quantum walk also contains a coin state, the amplitude at each vertex is uniformly distributed among its directions. That is,
\begin{align}
	\ket{s} = \frac{1}{\sqrt{N_1+N_2}} \bigg[ &\sum_{i\in X}\ket{i}\otimes\frac{1}{\sqrt{N_2}}\sum_{j\sim i} \ket{j} \label{eq:s} \\
	&+ \sum_{i\in Y}\ket{i}\otimes\frac{1}{\sqrt{N_1}}\sum_{j\sim i}\ket{j}\bigg]. \notag
\end{align}
When the graph is regular, applying the quantum walk $U_\text{walk} = SC$ (without the search query $Q$) to this initial state leaves it unchanged. That is, if $N_1 = N_2$, then $U_\text{walk} \ket{s} = \ket{s}$. This is what we would expect, as applying the quantum walk alone does not yield any information about the marked vertex since we did not query the oracle, so our initial equal guess over all the vertices is unchanged.

The complete bipartite graph, however, can be irregular. In this case, $\ket{s}$ is no longer stationary under the quantum walk. That is, when $N_1 \ne N_2$, then $U_\text{walk} \ket{s} \ne \ket{s}$. So the system evolves, even though we have not gained any information about which vertices may be marked. So in this paper, we also consider the following initial state, which is a uniform superposition over the edges:
\begin{equation}
	\label{eq:sigma}
	\ket{\sigma} = \frac{1}{\sqrt{2 N_1 N_2}} \left( \sum_{i = 1}^N \ket{i} \otimes \sum_{j \sim i} \ket{j} \right),
\end{equation}
Here, the probability of pointing along each edge is the same. This state has the property that $U_\text{walk} \ket{\sigma}=\ket{\sigma}$, which we desired. On the other hand, since vertices can have different numbers of edges, the probability of finding the particle at each vertex may not be uniform. A similar state was investigated for the continuous-time quantum walk in \cite{Wong19}.

In such spatial search problems \cite{CG2004}, the underlying graph is assumed to be known. This is akin to knowing the physical arrangement of data, such as a tape drive's data stored in a ribbon or a hard drive's data arranged in cylinders. Given this, it is possible to construct the initial states $\ket{s}$ and $\ket{\sigma}$ since they only require knowledge of the underlying graph; they can be prepared without knowledge of the marked vertices.

In the next Section, we explore search on the complete bipartite graph where the marked vertices are all in the same partite set, and we consider both starting states $\ket{s}$ and $\ket{\sigma}$. We find that for $\ket{s}$, the success probability at even and odd timesteps can vary greatly as the graph becomes more irregular. On the other hand, using $\ket{\sigma}$, the success probability is smoother and consistently reaches $1/2$. Following this, in Section III, we consider the case where the marked vertices lie in both partite sets. We will see that with either initial state, the success probability in a partite set only depends on the number of vertices (marked and unmarked) in that partite set. Again, $\ket{s}$ can cause the success probability to vary greatly from one timestep to the next, whereas $\ket{\sigma}$ yields a smoother evolution that consistently reaches a success probability of $1/2$ in each partite set. Note when the graph is regular, the two initial states are equivalent, and we show that search behaves differently from the complete graph \cite{Wong10}, which is Grover's algorithm formulated as a quantum walk. This is different from the continuous-time quantum walk, which asymptotically searches the regular complete bipartite graph \cite{Wong19} just like the complete graph \cite{Wong10}.


\section{Marked Vertices in One Set}

\begin{figure}
\begin{center}
	\includegraphics{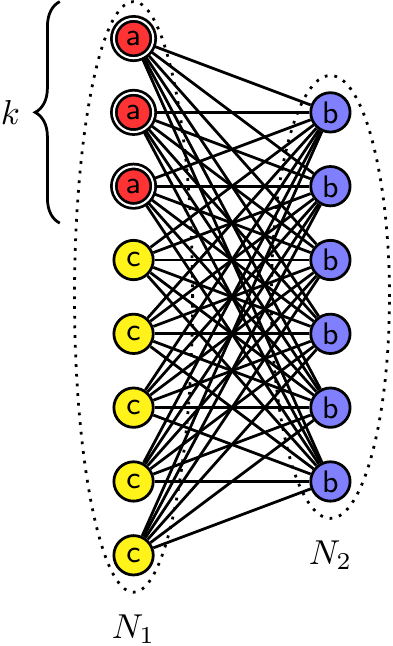}
	\caption{\label{fig:bipartite_leftmarked} The generalized complete bipartite graph containing $N_1$ vertices in set $X$ and $N_2$ vertices in set $Y$. There are $k$ marked vertices in set $X$, indicated by double circles. The vertices that evolve identically share the same color and label.}
\end{center}
\end{figure}

In this section, we consider the case where there are $k$ marked vertices, and they are all located in a single partite set, as shown in Fig.~\ref{fig:bipartite_leftmarked}. Without loss of generality, we take them to be in set $X$. With either initial state $\ket{s}$ or $\ket{\sigma}$, and evolution by $U$ \eqref{eq:U}, the system evolves in a four-dimensional (4D) subspace. This is because there are only three types of vertices, which we have labeled in Fig.~\ref{fig:bipartite_leftmarked}: the marked vertices labeled $a$, their adjacent vertices labeled $b$, and their nonadjacent vertices labeled $c$. A particle at an $a$ vertex can only point toward $b$ vertices, a particle at a $b$ vertex can point toward $a$ and/or $c$ vertices, and a particle at a $c$ vertex can only point toward $b$ vertices. Together, these yield the following orthonormal basis for the 4D subspace:
\begin{align*}
	\ket{ab} &= \frac{1}{\sqrt{k}}\sum_a\ket{a}\otimes\frac{1}{\sqrt{N_2}}\sum_b\ket{b}, \\
	\ket{ba} &= \frac{1}{\sqrt{N_2}}\sum_b\ket{b}\otimes\frac{1}{\sqrt{k}}\sum_a\ket{a}, \\
	\ket{bc} &= \frac{1}{\sqrt{N_2}}\sum_b\ket{b}\otimes\frac{1}{\sqrt{N_1-k}}\sum_c\ket{c}, \\
	\ket{cb} &= \frac{1}{\sqrt{N_1-k}}\sum_c\ket{c}\otimes\frac{1}{\sqrt{N_2}}\sum_b\ket{b}.
\end{align*}

Using (9) from \cite{Wong18}, we can write the quantum walk operator $U_\text{walk}$ in this 4D subspace. Furthermore, since only $\ket{ab}$ corresponds to a particle located at a marked vertex, the query is $Q = \text{diag}\{ \begin{matrix} -1 & 1 & 1 & 1 \end{matrix} \}$ in the 4D subspace. Together, the search operator in the $\{ \ket{ab}, \ket{ba}, \ket{bc}, \ket{cb} \}$ basis is
\begin{equation} 
	\label{eq:U_oneset}
	U = \begin{pmatrix}
		0 & -\cos\theta & \sin\theta & 0 \\
		-1 & 0 & 0 & 0 \\
		0 & 0 & 0 & 1 \\
		0 & \sin\theta & \cos\theta & 0 \\
	\end{pmatrix},
\end{equation}
where 
\begin{equation}
	\label{eq:theta}
	\cos\theta=1-\frac{2k}{N_1} \quad\text{and}\quad \sin\theta=\frac{2}{N_1}\sqrt{k(N_1-k)}.
\end{equation}

To find the evolution of the system for each starting state, we will need the eigenvectors and eigenvalues of $U$, which are
\begin{equation}
\begin{aligned}
	\ket{\psi_1} &= \frac{1}{2}\left[i,ie^{i\phi},-e^{i\phi},1\right]^\intercal ,\quad \lambda_1 = -e^{-i\phi}, \\
	\ket{\psi_2} &= \frac{1}{2}\left[i,-ie^{i\phi},e^{i\phi},1\right]^\intercal ,\quad \lambda_2 = e^{-i\phi}, \\
	\ket{\psi_3} &= \frac{1}{2}\left[-i,-ie^{-i\phi},-e^{-i\phi},1\right]^\intercal ,\quad \lambda_3 = -e^{i\phi}, \\
	\ket{\psi_4} &= \frac{1}{2}\left[-i,-ie^{-i\phi},e^{-i\phi},1\right]^\intercal ,\quad \lambda_4 = e^{i\phi},
\end{aligned} \label{eq:vecs_oneset}
\end{equation}
where $^\intercal$ denotes transpose, and
\begin{equation}
	\label{eq:phi}
	\phi = \frac{\theta}{2}.
\end{equation}

Now let us consider each initial state separately.


\subsection{Uniform Initial State Over Vertices}

\begin{figure*}
\begin{center}
        \subfloat[] {
		\includegraphics{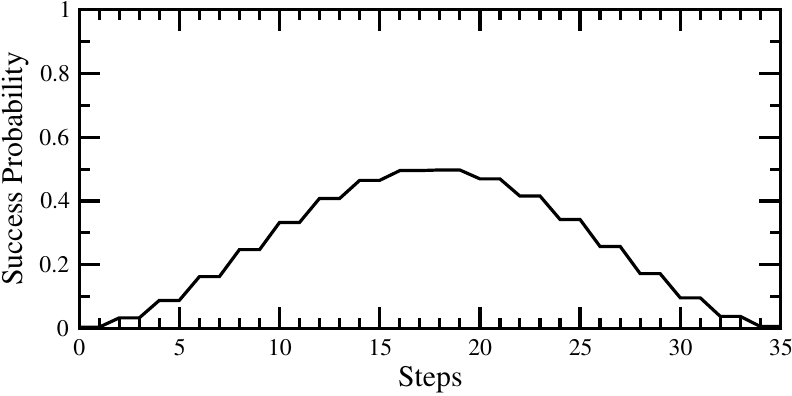}
                \label{fig:s_3_0_400_400}
        } \quad \quad
	\subfloat[] {
		\includegraphics{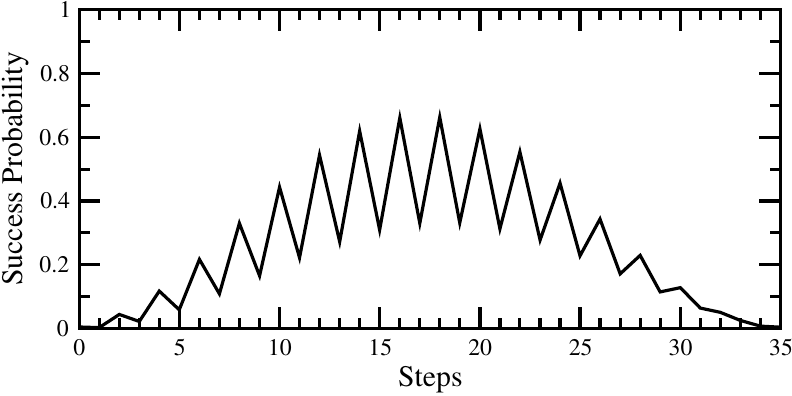}
                \label{fig:s_3_0_400_200}
        }

	\subfloat[] {
		\includegraphics{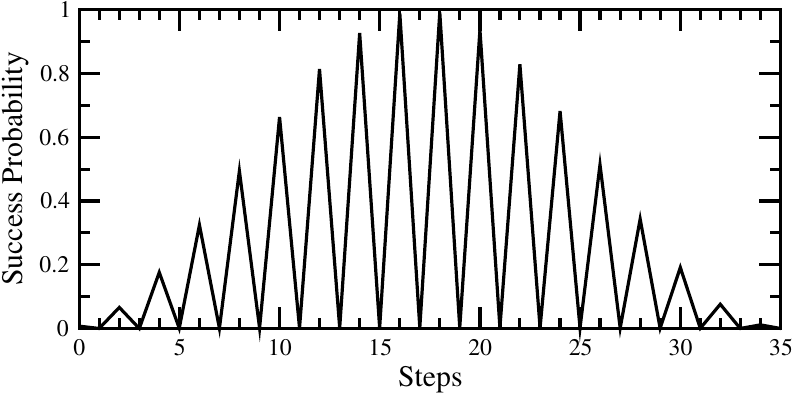}
                \label{fig:s_3_0_400_1}
        } \quad \quad
        \subfloat[] {
		\includegraphics{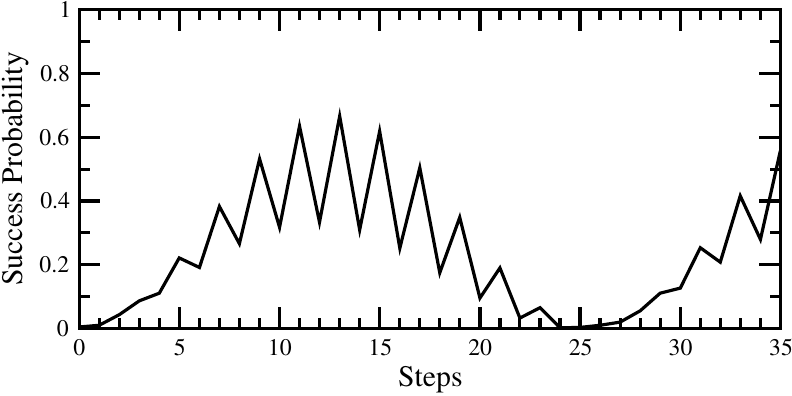}
                \label{fig:s_3_0_200_400}
        }

        \subfloat[] {
		\includegraphics{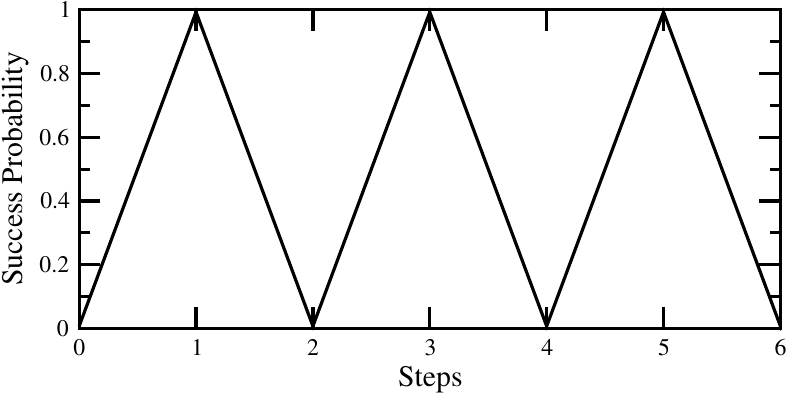}
                \label{fig:s_3_0_3_400}
        }
	\caption{\label{fig:s_3_0} Success probability for search on the complete bipartite graph from initial state $\ket{s}$ with $k=3$ marked vertices and (a) $N_1=N_2=400$ vertices, (b) $N_1=400$ vertices and $N_2=200$ vertices, (c) $N_1=400$ vertices and $N_2=1$ vertex, (d) $N_1=200$ vertices and $N_2=400$ vertices, and (e) $N_1=3$ vertices and $N_2=400$ vertices.}
\end{center}
\end{figure*}

Here, we consider the uniform state over the vertices $\ket{s}$ \eqref{eq:s} as the initial state. In the 4D subspace spanned by $\{ \ket{ab}, \ket{ba}, \ket{bc}, \ket{cb} \}$, it is
\begin{align*}
	\ket{s} 
		&= \frac{1}{\sqrt{N_1+N_2}} \bigg[ \sqrt{k} \ket{ab} + \sqrt{\frac{N_2}{N_1}k}\: \ket{ba}   \\
		&\quad\quad\quad\quad+ \sqrt{\frac{N_2}{N_1}(N_1-k)} \ket{bc} + \sqrt{N_1-k} \ket{cb}\bigg].
\end{align*}
The system evolves by repeatedly multiplying this 4D vector by $U$ \eqref{eq:U_oneset}, and the success probability at time $t$ is given by $p(t) = | \langle ab | U^t | s \rangle |^2$. Calculating this numerically, $p(t)$ is plotted in Fig.~\ref{fig:s_3_0} with $k=3$ marked vertices and with various values of $N_1$ and $N_2$. We see that when $N_1 = N_2$, as in Fig.~\ref{fig:s_3_0_400_400}, the success probability is somewhat smooth from one timestep to the next, reaching a maximum value of $1/2$. But as the ratio of $N_1$ to $N_2$ becomes greater, as in Figs.~\ref{fig:s_3_0_400_200} and \ref{fig:s_3_0_400_1}, the success probability from one timestep to the next becomes more and more jagged, with crests at even timesteps and troughs at odd timesteps. From Fig.~\ref{fig:s_3_0_400_1}, these crests and troughs can even alternate between (nearly) $0$ and $1$. This indicates that the runtime must be known precisely---measuring the system one timestep later or one timestep earlier can result in a success probability of zero. Furthermore, Figs.a-c have the same values of $N_1$ and $k$, so only $N_2$ is changing, yet they all peak in success probability after $18$ steps. Thus, their runtimes do not depend on $N_2$. Next, Figs.~\ref{fig:s_3_0_200_400} and \ref{fig:s_3_0_3_400} decrease the ratio of $N_1$ to $N_2$, and again the success probability is jagged, but now it crests at odd timesteps and troughs at even timesteps. In Fig.~\ref{fig:s_3_0_3_400}, all the vertices in set $X$ are marked, and the success probability roughly alternates between $0$ and $1$. Next, we prove these observations analytically.

\begin{table*}
\caption{\label{table:summary}Comparison of the two initial states, $\ket{s}$ and $\ket{\sigma}$, in terms of their asymptotic runtimes and maximum success probabilities for $N_1$ and $N_2$ large compared to the number of marked vertices.}
\begin{ruledtabular}
\def\arraystretch{1.7}
\begin{tabular}{ccc}
	Case & $\ket{s}$ & $\ket{\sigma}$  \\
	\hline
	$k_1 = k$, $k_2 = 0$ & $t_*=\frac{\pi}{2\sqrt{2}}\sqrt{\frac{N}{k}}$ & $t_*=\frac{\pi}{2\sqrt{2}}\sqrt{\frac{N}{k}}$ \\
	$N_1=N_2$ & $p_*=\frac{1}{2}$ & $p_*=\frac{1}{2}$ \\
	\hline
	$k_1=k$, $k_2=0$ & $t_*=\frac{\pi}{2}\sqrt{\frac{N_1}{k}}$ & $t_*=\frac{\pi}{2}\sqrt{\frac{N_1}{k}}$ \\
	$N_1>N_2$ & $p_*=\frac{N_1}{N_1+N_2}$ & $p_*=\frac{1}{2}$  \\
	\hline
	$k_1 = k$, $k_2=0$ & $t_*=\frac{\pi}{2}\sqrt{\frac{N_1}{k}}$ & $t_*=\frac{\pi}{2}\sqrt{\frac{N_1}{k}}$  \\
	$N_1 < N_2$ & $p_*=\frac{N_2}{N_1+N_2}$ & $p_*=\frac{1}{2}$ \\
	\hline
	\multirow{2}{*}{$\displaystyle \frac{k_1}{N_1} = \frac{k_2}{N_2}$} & $t_*=\frac{\pi}{2\sqrt{2}}\sqrt{\frac{N_1}{k_1}}$ & $t_*=\frac{\pi}{2\sqrt{2}}\sqrt{\frac{N_1}{k_1}}$ \\
	& $p_*=1$ & $p_*=1$ \\
	\hline
	$k_1=k_2$ & $t_{X*}=\frac{\pi}{2}\sqrt{\frac{N_1}{k}}$, $t_{Y*}=\frac{\pi}{2}\sqrt{\frac{N_2}{k}}$  & $t_{X*}=\frac{\pi}{2}\sqrt{\frac{N_1}{k}}$, $t_{Y*}=\frac{\pi}{2}\sqrt{\frac{N_2}{k}}$  \\
	$N_1 < N_2$ & $p_{X*}=\frac{N_2}{N_1+N_2}$, $p_{Y*}=\frac{N_2}{N_1+N_2}$ & $p_{X*}=\frac{1}{2}$, $p_{Y*}=\frac{1}{2}$ \\
	\hline
	$k_1<k_2$ & $t_{X*}=\frac{\pi}{2\sqrt{2}}\sqrt{\frac{N}{k_1}}$, $t_{Y*}=\frac{\pi}{2\sqrt{2}}\sqrt{\frac{N}{k_2}}$  & $t_{X*}=\frac{\pi}{2\sqrt{2}}\sqrt{\frac{N}{k_1}}$, $t_{Y*}=\frac{\pi}{2\sqrt{2}}\sqrt{\frac{N}{k_2}}$  \\
	$N_1 = N_2$ & $p_{X*}=\frac{1}{2}$, $p_{Y*}=\frac{1}{2}$ & $p_{X*}=\frac{1}{2}$, $p_{Y*}=\frac{1}{2}$ \\
	\hline
	$k_1<k_2$ & $t_{X*}=\frac{\pi}{2}\sqrt{\frac{N_1}{k_1}}$, $t_{Y*}=\frac{\pi}{2}\sqrt{\frac{N_2}{k_2}}$ & $t_{X*}=\frac{\pi}{2}\sqrt{\frac{N_1}{k_1}}$, $t_{Y*}=\frac{\pi}{2}\sqrt{\frac{N_2}{k_2}}$ \\
	$N_1 < N_2$ & $p_{X*}=\frac{N_2}{N_1+N_2}$, $p_{Y*}=\frac{N_2}{N_1+N_2}$ & $p_{X*}=\frac{1}{2}$, $p_{Y*}=\frac{1}{2}$ \\
	\hline
	$k_1<k_2$ & $t_{X*}=\frac{\pi}{2}\sqrt{\frac{N_1}{k_1}}$, $t_{Y*}=\frac{\pi}{2}\sqrt{\frac{N_2}{k_2}}$ & $t_{X*}=\frac{\pi}{2}\sqrt{\frac{N_1}{k_1}}$, $t_{Y*}=\frac{\pi}{2}\sqrt{\frac{N_2}{k_2}}$ \\
	$N_1 > N_2$ & $p_{X*}=\frac{N_1}{N_1+N_2}$, $p_{Y*}=\frac{N_1}{N_1+N_2}$ & $p_{X*}=\frac{1}{2}$, $p_{Y*}=\frac{1}{2}$ \\
\end{tabular}
\end{ruledtabular}
\end{table*}

To analytically determine the evolution, we express $\ket{s}$ as a linear combination of the eigenvectors of $U$ \eqref{eq:vecs_oneset}:
\[ \ket{s}=a\ket{\psi_1}+b\ket{\psi_2}+c\ket{\psi_3}+d\ket{\psi_4}, \]
where
\begin{align*}
	& a=\frac{\sqrt{N_1-k}-i\sqrt{k} - e^{-i\phi}\sqrt{\frac{N_2}{N_1}} \left( \sqrt{N_1-k} + i\sqrt{k} \right)}{2\sqrt{N_1+N_2}}, \\
	& b=\frac{\sqrt{N_1-k}-i\sqrt{k} + e^{-i\phi}\sqrt{\frac{N_2}{N_1}} \left( \sqrt{N_1-k} + i\sqrt{k} \right)}{2\sqrt{N_1+N_2}}, \\
	& c=\frac{\sqrt{N_1-k}+i\sqrt{k} - e^{i\phi}\sqrt{\frac{N_2}{N_1}}\left(\sqrt{N_1-k} - i\sqrt{k} \right)}{2\sqrt{N_1+N_2}}, \\
	& d=\frac{\sqrt{N_1-k}+i\sqrt{k} + e^{i\phi}\sqrt{\frac{N_2}{N_1}}\left(\sqrt{N_1-k} - i\sqrt{k} \right)}{2\sqrt{N_1+N_2}}.
\end{align*}
Applying the search operator pulls out eigenvalues, so the state of the system at time $t$ is
\[ U^t\ket{s} = a\lambda_1^t \ket{\psi_1} + b\lambda_2^t \ket{\psi_2} + c\lambda_3^t \ket{\psi_3} + d\lambda_4^t \ket{\psi_4}. \]
Multiplying on the left by $\bra{ab}$, substituting for the coefficients ($a$, $b$, $c$, $d$), substituting for the eigenvalues ($\lambda_i$'s), and squaring, the success probability at even and odd timesteps is
\begin{equation}
\label{eq:oneset_probtime}
\begin{aligned}
	p_\text{even} &= \frac{1}{N_1+N_2} \Big[ \sqrt{N_1 - k} \sin(\phi t) \\
		&\quad\quad\quad\quad+ \sqrt{k} \cos(\phi t) \Big]^2, \\
	p_\text{odd} &= \frac{N_2}{N_1(N_1+N_2)} \Big\{ \sqrt{N_1 - k} \sin[(1+t)\phi] \\
		&\quad\quad\quad\quad- \sqrt{k} \cos[(1+t)\phi] \Big\}^2.
\end{aligned}
\end{equation}
Together, these analytical results agree perfectly with the numerical simulations in Fig.~\ref{fig:s_3_0}.

Now the runtime of the algorithm is the time at which $p_\text{even}$ and $p_\text{odd}$ \eqref{eq:oneset_probtime} reach their first maxima. They are
\begin{equation}
\label{eq:oneset_times_exact}
\begin{aligned}
	t_\text{even} &= \frac{1}{\phi} \cos^{-1} \left( \sqrt{\frac{k}{N_1}} \right), \\
	t_\text{odd} &= t_\text{even} + 1 .
\end{aligned}
\end{equation}
So the maximum success probability at even and odd timesteps occur right after each other. Furthermore, recall from \eqref{eq:phi} and \eqref{eq:theta} that $\phi$ only depends on $N_1$ and $k$, so the runtime only depends on the number of marked and unmarked vertices in set $X$, in agreement with Figs.~\ref{fig:s_3_0}a\nobreakdash-c. Plugging these runtimes into \eqref{eq:oneset_probtime}, the corresponding maximum success probabilities are
\begin{equation}
\label{eq:oneset_probs}
\begin{gathered}
	p_\text{even} = \frac{N_1}{N_1+N_2}, \\
	p_\text{odd} = \frac{N_2}{N_1+N_2}.
\end{gathered}
\end{equation}

Using these formulas, we can prove additional behaviors from Fig.~\ref{fig:s_3_0}. First, when $N_1$ and $N_2$ are large, the runtimes \eqref{eq:oneset_times_exact} are asymptotically
\begin{equation}
	\label{eq:oneset_times}
	t_\text{even} = t_\text{odd} = \frac{\pi}{2} \sqrt{\frac{N_1}{k}}.
\end{equation}
Now when $N_1 = N_2 = N/2$, as in Fig.~\ref{fig:s_3_0_400_400}, then for large $N$, the runtimes \eqref{eq:oneset_times} are
\[ t_* = \frac{\pi}{2\sqrt{2}} \sqrt{\frac{N}{k}}, \]
and the corresponding success probabilities \eqref{eq:oneset_probs} are
\[ p_* = \frac{1}{2}. \]
This is summarized in the first row of Table~\ref{table:summary}, and it agrees with Fig.~\ref{fig:s_3_0_400_400}, where the success probability reaches $1/2$ at time $(\pi/2\sqrt{2})\sqrt{800/3} \approx 18$. Note search on the complete graph, which is Grover's algorithm formulated as a quantum walk, also reaches its maximum success probability at time $(\pi/2\sqrt{2})\sqrt{N/k}$ \cite{Wong10}. But its maximum success probability is $1/2$ when $k = 1$ and $4k(k-1)/(2k-1)^2$ when $k \ge 2$ \cite{Wong10}, so it only matches the regular complete bipartite graph for a single marked vertex. This result differs from the continuous-time quantum walk, where the complete graph and the regular complete bipartite graph behave the same way for large $N_1$ and $N_2$, even with multiple marked vertices \cite{Wong19}.

\begin{figure*}
\begin{center}
        \subfloat[] {
		\includegraphics{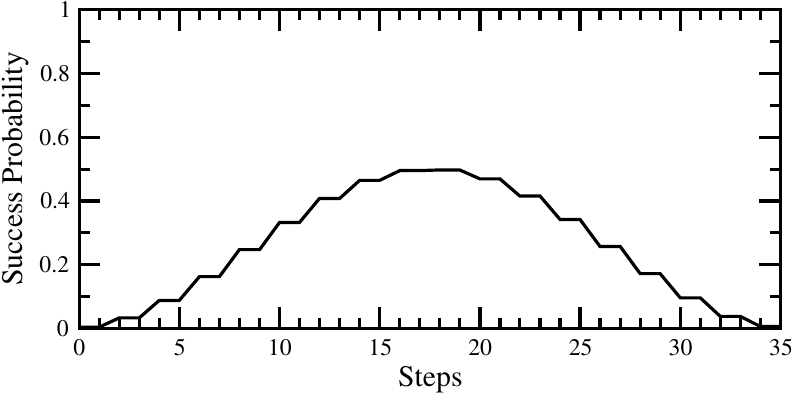}
                \label{fig:sigma_3_0_400_400}
        } \quad \quad
	\subfloat[] {
		\includegraphics{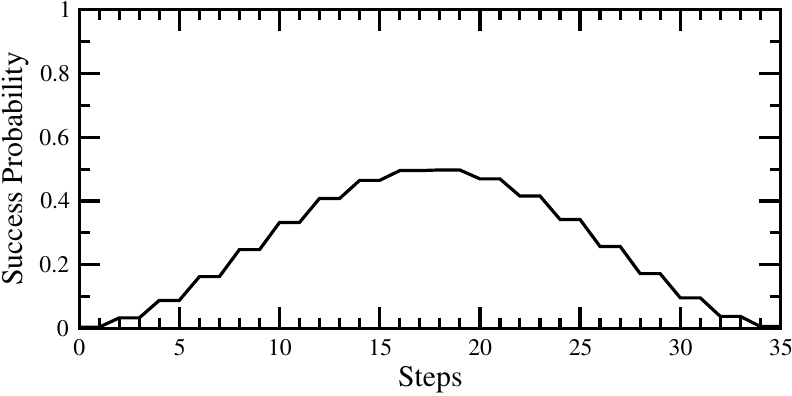}
                \label{fig:sigma_3_0_400_200}
        }

	\subfloat[] {
		\includegraphics{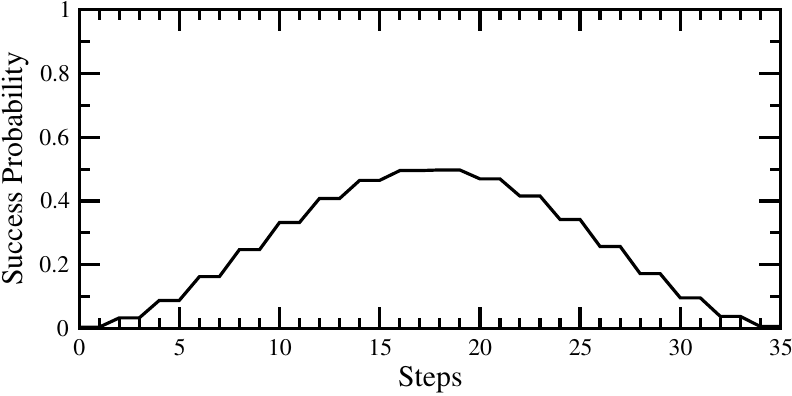}
                \label{fig:sigma_3_0_400_1}
        } \quad \quad
        \subfloat[] {
		\includegraphics{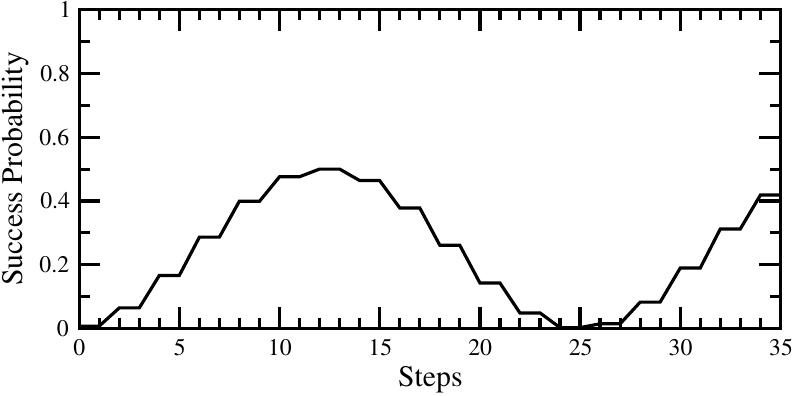}
                \label{fig:sigma_3_0_200_400}
        }

        \subfloat[] {
		\includegraphics{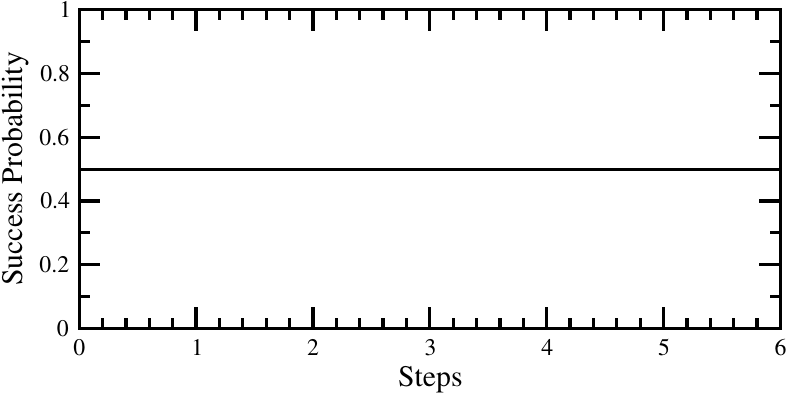}
                \label{fig:sigma_3_0_3_400}
        }

	\caption{\label{fig:sigma_3_0} Success probability for search on the complete bipartite graph from initial state $\ket{\sigma}$ with $k=3$ marked vertices and (a) $N_1=N_2=400$ vertices, (b) $N_1=400$ vertices and $N_2=200$ vertices, (c) $N_1=400$ vertices and $N_2=1$ vertex, (d) $N_1=200$ vertices and $N_2=400$ vertices, and (e) $N_1=3$ vertices and $N_2=400$ vertices.}
\end{center}
\end{figure*}

Next, when $N_1 > N_2$, as in Figs.~\ref{fig:s_3_0_400_200} and \ref{fig:s_3_0_400_1}, the success probability reaches a maximum value of $p_\text{even} = N_1 / (N_1 + N_2)$ and $p_\text{odd} = N_2 / (N_1 + N_2)$ at respective runtimes $t_\text{even}$ and $t_\text{odd}$ \eqref{eq:oneset_times}. These results agree numerically with Fig.~\ref{fig:s_3_0_400_200}, where the success probability reaches $400 / (400 + 200) \approx 0.667$ and $200 / (400 + 200) \approx 0.333$ at times $(\pi/2) \sqrt{400/3} = 18$ and $19$, respectively. Similarly, in Fig.~\ref{fig:s_3_0_400_1}, the success probability reaches $400 / (400 + 1) \approx 0.998$ and $1 / (400 + 1) \approx 0.002$ at times $(\pi/2) \sqrt{400/3} \approx 18$ and $19$, respectively. Since $N_1 > N_2$, $p_\text{even} > p_\text{odd}$, so we use the even success probability in the second row of Table~\ref{table:summary}.

Similarly, when $N_1 < N_2$, as in Figs.~\ref{fig:s_3_0_200_400} and \ref{fig:s_3_0_3_400}, then $p_\text{even} = N_1 / (N_1 + N_2) < p_\text{odd} = N_2 / (N_1 + N_2)$. Using $p_\text{odd}$, we get the third row of Table~\ref{table:summary}.

As these cases demonstrate, large differences between $N_1$ and $N_2$ create large differences in the success probability at even and odd steps, and therefore, it becomes important to account for this when measuring the system.


\subsection{Uniform Initial State Over Edges}

Now we instead consider the initial state $\ket{\sigma}$ \eqref{eq:sigma} that is a uniform superposition over the edges. In the 4D subspace spanned by $\{ \ket{ab}, \ket{ba}, \ket{bc}, \ket{cb} \}$, it is
\begin{align*}
	\ket{\sigma}
		&= \frac{1}{\sqrt{2N_1N_2}} \Big[ \sqrt{kN_2} \ket{ab} + \sqrt{kN_2} \ket{ba} \\
		&\quad + \sqrt{N_2(N_1-k)} \ket{bc} + \sqrt{N_2(N_1-k)} \ket{cb} \Big].
\end{align*}
Then the success probability at time $t$ is $p(t) = | \langle ab | U^t | \sigma \rangle |^2$. This is plotted in Fig.~\ref{fig:sigma_3_0} with the same choices of $k$, $N_1$, and $N_2$ as Fig.~\ref{fig:s_3_0}, allowing for a direct comparison between the two initial states. In Figs.~\ref{fig:s_3_0_400_400} and \ref{fig:sigma_3_0_400_400}, the graph is regular, so $\ket{s}$ and $\ket{\sigma}$ are equal, and we get the same evolution for both initial states. For the remaining subfigures, the graph is irregular, and we get a different evolutions with $\ket{\sigma}$ from $\ket{s}$. First, the evolution with $\ket{\sigma}$ is much smoother compared to the jaggedness with $\ket{s}$. Second, the success probability consistently reaches a maximum of 1/2. Third, the evolutions in Figs.~\ref{fig:sigma_3_0}a-c are identical, so if the marked vertices are in set $X$, the evolution does not depend on the number of vertices in $Y$. If we change the number of vertices in set $X$, however, then the evolution does change, as shown in Figs.~\ref{fig:sigma_3_0}d-e.

To prove this analytically, we again express the initial state as a linear combination of the eigenvectors $\eqref{eq:vecs_oneset}$ of the search operator $\eqref{eq:U_oneset}$:
\[ \ket{\sigma} = a'\ket{\psi_1} + b'\ket{\psi_2} + c'\ket{\psi_3} + d'\ket{\psi_4}, \]
where
\begin{align*}
	& a' = \frac{\sqrt{N_1-k} - i\sqrt{k} - e^{-i\phi} \left( \sqrt{N_1-k} + i\sqrt{k} \right)}{2\sqrt{2N_1}}, \\
	& b' = \frac{\sqrt{N_1-k} - i\sqrt{k} + e^{-i\phi} \left( \sqrt{N_1-k} + i\sqrt{k} \right)}{2\sqrt{2N_1}}, \\
	& c' = \frac{\sqrt{N_1-k} + i\sqrt{k} - e^{i\phi} \left( \sqrt{N_1-k} - i\sqrt{k} \right)}{2\sqrt{2N_1}}, \\
	& d' = \frac{\sqrt{N_1-k} + i\sqrt{k} + e^{i\phi} \left( \sqrt{N_1-k} - i\sqrt{k} \right)}{2\sqrt{2N_1}}.
\end{align*}
Then the success probability at time $t$ is given by $|\bra{ab}U^t\ket{\sigma}|^2$, and breaking it into even and odd timesteps, we get
\begin{align*}
	p_\text{even}(t) &= \left[\frac{\sqrt{N_1-k}\sin(\phi t)+\sqrt{k}\cos(\phi t)}{\sqrt{2 N_1}}\right]^2, \\
	p_\text{odd}(t) &= \left[\frac{\sqrt{N_1-k}\sin(\phi (t+1))-\sqrt{k}\cos(\phi (t+1))}{\sqrt{2 N_1}}\right]^2.
\end{align*}
The runtime of the algorithm is the time at which $p(t)$ reaches its first maximum, which at even and odd timesteps is
\begin{align*}
	t_{\text{even}} &= \frac{1}{\phi} \cos^{-1}\left(\sqrt{\frac{k}{N_1}}\right), \\
	t_{\text{odd}} &= t_\text{even} + 1.
\end{align*}
These are the same runtimes as $\ket{s}$ from \eqref{eq:oneset_times_exact}, so they have the same asymptotic forms as \eqref{eq:oneset_times}. Plugging into $p(t)$, however, the maximum success probability is now
\[ p_\text{even} = p_\text{odd} = \frac{1}{2}. \]
So the maximum success probability is always $1/2$, no matter which values $N_1$ and $N_2$ take. Even in the extreme case where there are only marked vertices in one set, the probability is always $1/2$, as shown in Fig.~\ref{fig:sigma_3_0_3_400} instead of alternating between zero and one as seen in Fig.~\ref{fig:s_3_0_3_400}. These results are summarized in the first three rows of Table~\ref{table:summary}. Thus, if the graph is regular, the two initial states are identical, but if the graph is irregular, $\ket{s}$ yields a better success probability than $\ket{\sigma}$, although one must be careful to measure at an even timestep when $N_1 > N_2$ and an odd timestep when $N_1 < N_2$.


\section{Marked Vertices in Both Sets}

\begin{figure}
\begin{center}
	\includegraphics{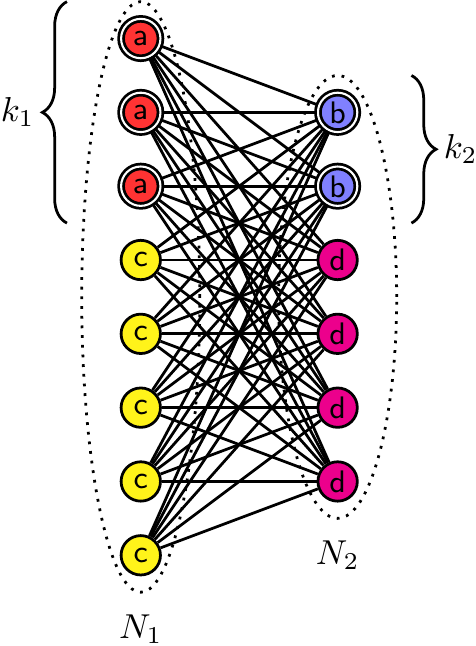}
	\caption{\label{fig:bipartite_bothmarked} The generalized complete bipartite graph containing $N_1$ vertices in set $X$ and $N_2$ vertices in set $Y$. There are $k_1$ marked vertices in set $X$ and $k_2$ marked vertices in set $Y$, indicated by double circles. The vertices that evolve identically share the same color and label.}
\end{center}
\end{figure}

Now we move on to the case where there are $k_1$ marked vertices in partite set $X$ and $k_2$ marked vertices in partite set $Y$, as shown in Fig.~\ref{fig:bipartite_bothmarked}. Then there are four types of vertices, and vertices can point to both marked and unmarked vertices in the other set. This leads to an 8D subspace spanned by the orthonormal basis states
\begin{align*}
	& \ket{ab}=\frac{1}{\sqrt{k_1}}\sum_a\ket{a}\otimes\frac{1}{\sqrt{k_2}}\sum_b\ket{b}, \\
	& \ket{ad}=\frac{1}{\sqrt{k_1}}\sum_a\ket{a}\otimes\frac{1}{\sqrt{N_2-k_2}}\sum_d\ket{d}, \\
	& \ket{ba}=\frac{1}{\sqrt{k_2}}\sum_b\ket{b}\otimes\frac{1}{\sqrt{k_1}}\sum_a\ket{a}, \\
	& \ket{bc}=\frac{1}{\sqrt{k_2}}\sum_b\ket{b}\otimes\frac{1}{\sqrt{N_1-k_1}}\sum_c\ket{c}, \\
	& \ket{cb}=\frac{1}{\sqrt{N_1-k_1}}\sum_c\ket{c}\otimes\frac{1}{\sqrt{k_2}}\sum_b\ket{b}, \\
	& \ket{cd}=\frac{1}{\sqrt{N_1-k_1}}\sum_c\ket{c}\otimes\frac{1}{\sqrt{N_2-k_2}}\sum_d\ket{d}, \\
	& \ket{da}=\frac{1}{\sqrt{N_2-k_2}}\sum_d\ket{d}\otimes\frac{1}{\sqrt{k_1}}\sum_a\ket{a}, \\
	& \ket{dc}=\frac{1}{\sqrt{N_2-k_2}}\sum_d\ket{d}\otimes\frac{1}{\sqrt{N_1-k_1}}\sum_c\ket{c}.
\end{align*}
In this basis, the quantum walk operator $U_\text{walk} = SC$ can be obtained using Eq.~[9] from \cite{Wong18}, and since both $a$ and $b$ vertices are marked, the oracle is $Q = \text{diag}\{ -1, -1, -1, -1, 1, 1, 1, 1 \}$ in this basis. Combining them, the search operator $U = U_\text{walk}Q$ \eqref{eq:U} is
\begin{widetext}
\begin{equation}
	U = \begin{pmatrix}
		0 & 0 & 1-\frac{2k_1}{N_1} & \frac{-2}{N_1}\sqrt{k_1N_{k_1}} & 0 & 0 & 0 & 0 \\
		0 & 0 & 0 & 0 & 0 & 0 & \frac{2k_1}{N_1}-1 & \frac{2}{N_1}\sqrt{k_1N_{k_1}} \\
		1-\frac{2k_2}{N_2} & \frac{-2}{N_2}\sqrt{k_2N_{k_2}} & 0 & 0 & 0 & 0 & 0 & 0 \\
		0 & 0 & 0 & 0 & \frac{2k_2}{N_2}-1 & \frac{2}{N_2}\sqrt{k_2N_{k_2}} & 0 & 0 \\
		0 & 0 & \frac{-2}{N_1}\sqrt{k_1N_{k_1}} & \frac{2k_1}{N_1}-1 & 0 & 0 & 0 & 0 \\
		0 & 0 & 0 & 0 & 0 & 0 & \frac{2}{N_1}\sqrt{k_1N_{k_1}} & 1-\frac{2k_1}{N_1} \\
		\frac{-2}{N_2}\sqrt{k_2N_{k_2}} & \frac{2k_2}{N_2}-1 & 0 & 0 & 0 & 0 & 0 & 0 \\
		0 & 0 & 0 & 0 & \frac{2}{N_2}\sqrt{k_2N_{k_2}} & 1-\frac{2k_2}{N_2} & 0 & 0 \\
	\end{pmatrix},
	\label{eq:U_bothmarked}
\end{equation}
\end{widetext}
where $N_{k_1}=N_1-k_1$ and $N_{k_2}=N_2-k_2$.

In order to find the evolution of the system, we want to find the eigenvalues and eigenvectors of $U$ and express the initial state as a linear combination of these eigenvectors. The exact eigenvectors are complicated and make it difficult to discern the behavior of the algorithm, so we assume that $k_1 = o(N_1)$ and $k_2 = o(N_2)$, otherwise we can classically sample for a marked vertex in each set in constant time, eliminating the need for a quantum search algorithm. With this assumption, the eigenvectors and eigenvalues of $U$ are asymptotically
\begin{equation}
	\begin{aligned}
		& \ket{\psi_1} = \frac{1}{2\sqrt{2}} \left[ 1, i, 1, -i, i, -1, -i, 0 \right]^\intercal, \quad \lambda_1=e^{-i\alpha}, \\
		& \ket{\psi_2} = \frac{1}{2\sqrt{2}} \left[ 1, -i, 1, i, -i, -1, i, 0 \right]^\intercal, \quad \lambda_2=e^{i\alpha}, \\
		& \ket{\psi_3} = \frac{1}{2\sqrt{2}} \left[ 1, i, 1, i, -i, 1, -i, 0 \right]^\intercal, \quad \lambda_3=e^{-i\beta}, \\
		& \ket{\psi_4} = \frac{1}{2\sqrt{2}} \left[  1, -i, 1, -i, i, 1, i, 0 \right]^\intercal, \quad \lambda_4=e^{i\beta}, \\
		& \ket{\psi_5} = \frac{1}{2\sqrt{2}} \left[ -1, i, 1, i, i, 1, i, -1 \right]^\intercal, \quad \lambda_5=-e^{i\alpha}, \\
		& \ket{\psi_6} = \frac{1}{2\sqrt{2}} \left[ -1, -i, 1, -i, -i, 1, -i, -1 \right]^\intercal, \quad \lambda_6=-e^{-i\alpha}, \\
		& \ket{\psi_7} = \frac{1}{2\sqrt{2}} \left[ -1, i, 1, -i, -i, -1, i, 1 \right]^\intercal, \quad \lambda_7=-e^{i\beta}, \\
		& \ket{\psi_8} = \frac{1}{2\sqrt{2}} \left[ -1, -i, 1, i, i, -1, -i, 1 \right]^\intercal, \quad \lambda_8=-e^{-i\beta}, \\
	\end{aligned}
	\label{eq:vecs_bothmarked}
\end{equation}
where
\[ \sin\alpha=\sqrt{\frac{k_2}{N_2}}+\sqrt{\frac{k_1}{N_1}} \quad\text{and}\quad \sin\beta=\sqrt{\frac{k_2}{N_2}}-\sqrt{\frac{k_1}{N_1}}. \]
The details of this derivation are given in Appendix~\ref{sec:appendix}. Next, let us consider search with each of the two starting states, beginning with $\ket{s}$.

\subsection{Uniform Initial State Over Vertices}

\begin{figure}
\begin{center}
        \subfloat[] {
                \includegraphics{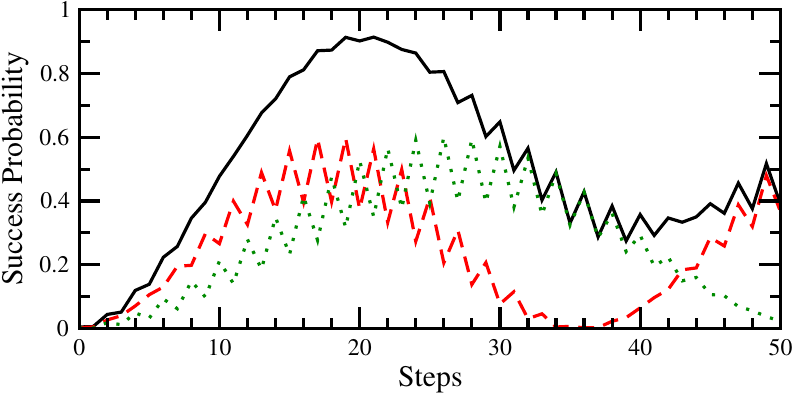}
                \label{fig:s_3_2_400_600}
        }

	\subfloat[] {
                \includegraphics{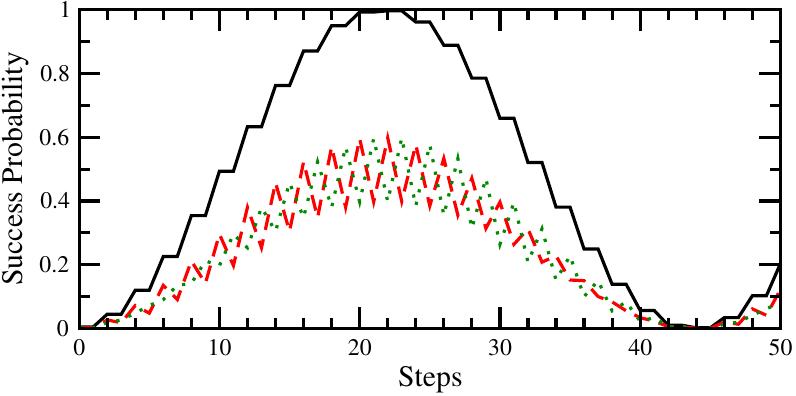}
                \label{fig:s_3_2_600_400}
        }

	\subfloat[] {
                \includegraphics{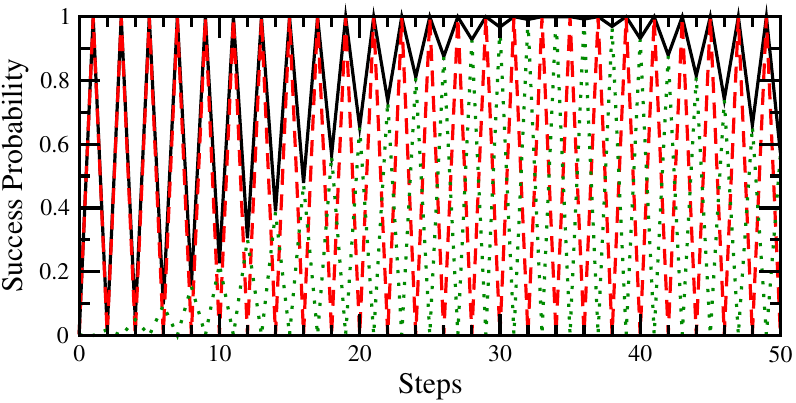}
                \label{fig:s_3_2_3_997}
	}
	\caption{\label{fig:s_3_2} Success probability for search on the complete bipartite graph from initial state $\ket{s}$ with $k_1=3$ marked vertices in set $X$, $k_2=2$ marked vertices in set $Y$, and (a) $N_1=400$ and $N_2=600$ vertices, (b) $N_1=600$ and $N_2=400$ vertices, and (c) $N_1=3$ and $N_2=997$ vertices. The solid black curve is the total probability at all the marked vertices, the dashed red curve is the probability at marked vertices in set $X$, and the dotted green curve is the probability at the marked vertices in set $Y$.}
\end{center}
\end{figure}

In the 8D basis, the typical initial state $\ket{s}$ \eqref{eq:s}, which is a uniform superposition over the vertices, is
\begin{align*}
	\ket{s}
		&= \frac{1}{\sqrt{N_1+N_2}} \bigg[ \sqrt{\frac{k_1k_2}{N_2}}\ket{ab}+\sqrt{\frac{k_1(N_2-k_2)}{N_2}}\ket{ad} \\
		&\quad+ \sqrt{\frac{k_1k_2}{N_1}}\ket{ba} + \sqrt{\frac{k_2(N_1-k_1)}{N_1}}\ket{bc} \\
		&\quad+ \sqrt{\frac{k_2(N_1-k_1)}{N_2}}\ket{cb} + \sqrt{\frac{(N_1-k_1)(N_2-k_2)}{N_2}}\ket{cd} \\
		&\quad+ \sqrt{\frac{k_1(N_2-k_2)}{N_1}}\ket{da} + \sqrt{\frac{(N_1-k_1)(N_2-k_2)}{N_1}}\ket{dc}\bigg].
\end{align*}
This exactly evolves by repeated applications of $U$ \eqref{eq:U_bothmarked}, and at time $t$, the success probability in set $X$ is $p_X(t) = |\langle ab | U^t | s \rangle|^2 + |\langle ad | U^t | s \rangle|^2$, and the success probability in set $Y$ is $p_Y(t) = |\langle ba | U^t | s \rangle|^2 + |\langle bc | U^t | s \rangle|^2$, so the total success probability is $p(t) = p_X(t) + p_Y(t)$. These are plotted in Fig.~\ref{fig:s_3_2} with $k_1 = 3$, $k_2 = 2$, and various values of $N_1$ and $N_2$. We see in Fig.~\ref{fig:s_3_2_400_600} that the peak success probability in $X$ and $Y$ do not coincide, so the total success probability does not reach $1$. In Fig.~\ref{fig:s_3_2_600_400}, however, they do align, and the success probability does reach $1$. Finally, in Fig.~\ref{fig:s_3_2_3_997}, all the vertices in set $X$ are marked, and the graph is highly irregular. The total success probability is jagged, roughly $1$ at odd timesteps, and gradually increasing at even timesteps until it is also $1$.

Now let us investigate these results analytically for large $N_1$ and $N_2$. Then we can explain Figs.~\ref{fig:s_3_2_400_600} and \ref{fig:s_3_2_600_400}, but unfortunately, Fig.~\ref{fig:s_3_2_3_997} has $k_1 = N_1$, so it does not satisfy $N_1$ large compared to $k_1$. For large $N_1$ and $N_2$, the initial state is asymptotically
\[ \ket{s} = \frac{1}{\sqrt{N_1 + N_2}} \left( \sqrt{N_1} \ket{cd} + \sqrt{N_2} \ket{dc} \right). \]
Expressing this as a linear combination of the asymptotic eigenvectors of $U$ \eqref{eq:vecs_bothmarked},
\begin{align*}
	\ket{s} 
		&= a\ket{\psi_1}+b\ket{\psi_2}+c\ket{\psi_3}+d\ket{\psi_4} \\
		&\quad+ e\ket{\psi_5}+f\ket{\psi_6}+g\ket{\psi_7}+h\ket{\psi_8},
\end{align*}
where 
\begin{align*}
	& a = b= \frac{-\sqrt{N_1}-\sqrt{N_2}}{2\sqrt{2}\sqrt{N_1+N_2}}, \\
	& c = d= \frac{\sqrt{N_1}+\sqrt{N_2}}{2\sqrt{2}\sqrt{N_1+N_2}}, \\
	& e = f= \frac{\sqrt{N_1}-\sqrt{N_2}}{2\sqrt{2}\sqrt{N_1+N_2}}, \\
	& g = h= \frac{-\sqrt{N_1}+\sqrt{N_2}}{2\sqrt{2}\sqrt{N_1+N_2}}.
\end{align*}
Applying $U$ multiplies each eigenvector by its eigenvalue, and adding the squares of the amplitudes in $\ket{ab}$, $\ket{ad}$, $\ket{ba}$, and $\ket{bc}$, the success probability at time $t$ is
\[ p(t) = \begin{cases} 
	\frac{N_1\sin^2\left[(\alpha-\beta)t/2\right] + N_2\sin^2\left[(\alpha+\beta)t/2\right]}{N_1+N_2}, & t\text{ even,} \\
	\frac{N_2\sin^2\left[(\alpha-\beta)t/2\right] + N_1\sin^2\left[(\alpha+\beta)t/2\right]}{N_1+N_2}, & t\text{ odd.} \\
\end{cases} \]
To find the runtime and maximum success probability, we take the derivative of $p(t)$, set it equal to zero, and solve for the runtime. This does not have a closed-form solution, however. So instead, we evaluate the success probability in each partite set independently. The success probability in $X$ is
\[ p_X(t) = \begin{cases} 
	\frac{N_1\sin^2\left[(\alpha-\beta)t/2\right]}{N_1+N_2}, & t \text{ even}, \\
	\frac{N_2\sin^2\left[(\alpha-\beta)t/2\right]}{N_1+N_2}, & t \text{ odd},
\end{cases} \]
and the success probability in set $Y$ is
\[ p_Y(t) = \begin{cases} 
	\frac{N_2\sin^2\left[(\alpha+\beta)t/2\right]}{N_1+N_2}, & t \text{ even}, \\
	\frac{N_1\sin^2\left[(\alpha+\beta)t/2\right]}{N_1+N_2}, & t \text{ odd.} 
\end{cases} \]
These reach their respective maxima at
\[ t_{X*} = \frac{\pi}{\alpha-\beta}, \quad t_{Y*} = \frac{\pi}{\alpha+\beta}. \]
When we consider large values of $N_1$ and $N_2$, $\alpha$ and $\beta$ are small, and so $\sin\alpha\approx\alpha$ and $\sin\beta\approx\beta$. Substituting these values into $t_X$ and $t_Y$, they become
\[ t_{X*} = \frac{\pi}{2}\sqrt{\frac{N_1}{k_1}}, \quad \quad t_{Y*} = \frac{\pi}{2}\sqrt{\frac{N_2}{k_2}}. \]
This shows that the probability in each set evolves independently of the other, so the runtime of set $X$ only depends on the number of vertices $N_1$ and marked vertices $k_1$, and the runtime of set $Y$ only depends on the number of vertices $N_2$ and marked vertices $k_2$.

The maximum success probability can be found by substituting the runtimes into $p_X$ and $p_Y$, which yields
\[ p_{X*} = \begin{cases} 
	\frac{N_1}{N_1+N_2}, & t \text{ even}, \\
	\frac{N_2}{N_1+N_2}, & t \text{ odd}, 
\end{cases} \]
and
\[ p_{Y*} = \begin{cases} 
	\frac{N_2}{N_1+N_2}, & t \text{ even}, \\
	\frac{N_1}{N_1+N_2}, & t \text{ odd}.
\end{cases} \]

We can apply these results to specific cases. First, if the ratios $k_1 / N_1$ and $k_2 / N_2$ are equal, then the runtimes $t_{X*}$ and $t_{Y*}$ are equal. So the success probability in each set peaks at the same time, and the total success probability reaches a peak of $p_* = p_{X*} + p_{Y*} = 1$. This proves the behavior exhibited in Fig.~\ref{fig:s_3_2_600_400}, and it is summarized in the fourth row of Table~\ref{table:summary}. A specific case of $k_1 / N_1 = k_2 / N_2$ is when $k_1 = k_2 = k/2$ and $N_1 = N_2 = N/2$. In this case, the runtime is $(\pi/2\sqrt{2}) \sqrt{N/k}$, which is the same runtime as search on the complete graph \cite{Wong10}. Although the runtimes are the same, the success probabilities differ. Here, we have a success probability of $p_* = 1$, but the complete graph's success probability is $1/2$ when $k = 1$ and $4k(k-1)/(2k-1)^2$ when $k \ge 2$ \cite{Wong10}, so they only converge as $k$ grows. Thus, while search on the regular complete bipartite graph and the complete graph have the same runtime when $k_1 = k_2$ and $N_1 = N_2$, their success probabilities differ, and so search on them by coined quantum walk is different. This is a stark contrast to the continuous-time quantum walk, where search on the regular complete bipartite graph \cite{Wong19} behaves exactly the same way as search on the complete graph for any number of marked vertices, as long as $k_1 = o(N_1)$ and $k_2 = o(N_2)$.

Next, if $k_1 = k_2$ but $N_1 \neq N_2$, then the success probability in each partite set will not peak at the same time. Without loss of generality, we can assume $N_1 < N_2$. If not, we can exchange the sets $X$ and $Y$, and then it will be true. Then the maximum success probability in each set is $N_2 / (N_1 + N_2)$, and this is summarized in the fifth row of Table~\ref{table:summary}.

Now if the number of marked vertices in each set is different, we can assume $k_1 < k_2$ or exchange the sets $X$ and $Y$. The behavior now depends on the relationship between $N_1$ and $N_2$. If $N_1 = N_2$, then $t_{X*} \ne t_{Y*}$, but $p_{X*} = p_{Y*} = 1/2$. This is summarized in the sixth row of Table~\ref{table:summary}. If instead $N_1 < N_2$, then $p_{X*} = p_{Y*} = N_2/(N_1 + N_2)$, and this is illustrated in Fig.~\ref{fig:s_3_2_400_600} and summarized in the seventh row of Table~\ref{table:summary}. Finally, when $N_1 > N_2$, then $p_{X*} = p_{Y*} = N_1/(N_1 + N_2)$, and this is summarized in the eight row of Table~\ref{table:summary}.


\subsection{Uniform Initial State Over Edges}

\begin{figure}
\begin{center}
        \subfloat[] {
		\includegraphics{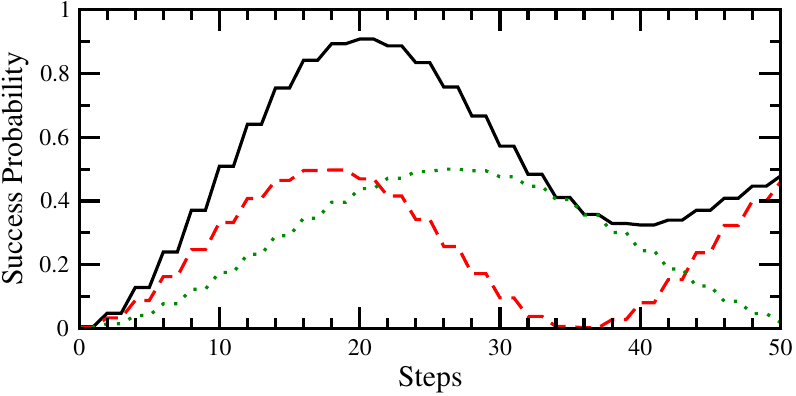}
                \label{fig:sigma_3_2_400_600}
        } \quad \quad
	\subfloat[] {
                \includegraphics{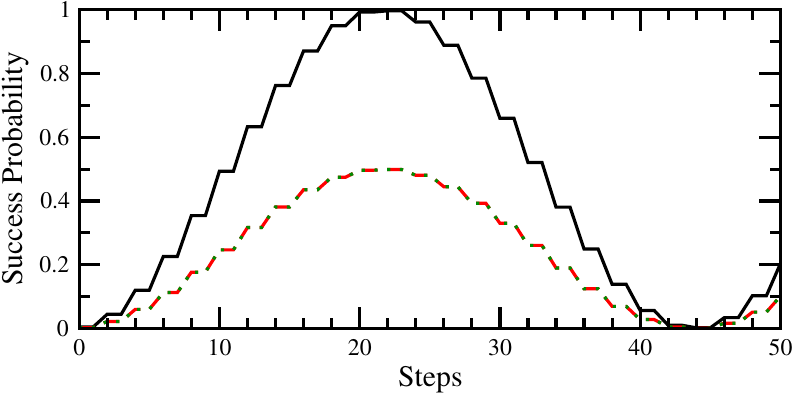}
                \label{fig:sigma_3_2_600_400}
        } \quad \quad
	\subfloat[] {
                \includegraphics{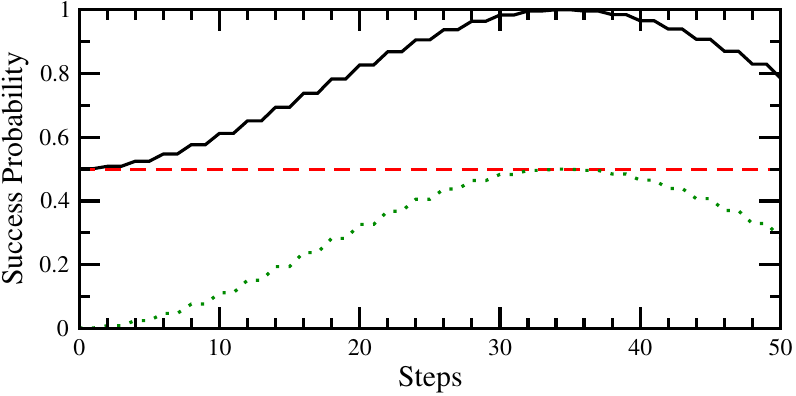}
                \label{fig:sigma_3_2_3_997}
	}
	\caption{\label{fig:sigma_3_2} Success probability for search on the complete bipartite graph from initial state $\ket{\sigma}$ with $k_1=3$ marked vertices in set $X$, $k_2=2$ marked vertices in set $Y$, and (a) $N_1=400$ and $N_2=600$ vertices, (b) $N_1=600$ and $N_2=400$ vertices, and (c) $N_1=3$ and $N_2=997$ vertices. The solid black curve is the total probability at all the marked vertices, the dashed red curve is the probability at marked vertices in set $X$, and the dotted green curve is the probability at the marked vertices in set $Y$.}
\end{center}
\end{figure}

Finally, the initial superposition over the edges $\ket{\sigma}$ \eqref{eq:sigma} is, in the $\{ \ket{ab}, \ket{ad}, \ket{ba}, \ket{bc}, \ket{cb}, \ket{cd}, \ket{da}, \ket{dc} \}$ basis,
\begin{align*}
	\ket{\sigma} &= \frac{1}{\sqrt{2N_1N_2}} \bigg[ \sqrt{k_1k_2}\ket{ab}+\sqrt{k_1(N_2-k_2)}\ket{ad} \\
		&\quad+ \sqrt{k_1k_2}\ket{ba} + \sqrt{k_2(N_1-k_1)}\ket{bc} \\
		&\quad+ \sqrt{k_2(N_1-k_1)}\ket{cb} + \sqrt{(N_1-k_1)(N_2-k_2)}\ket{cd} \\
		&\quad+ \sqrt{k_1(N_2-k_2)}\ket{da} + \sqrt{(N_1-k_1)(N_2-k_2)}\ket{dc}\bigg].
\end{align*}
As before, this evolves by repeated application of $U$ \eqref{eq:U_bothmarked}, and the success probability at time $t$ in set $X$ is $p_X(t) = |\langle ab | U^t | \sigma \rangle|^2 + |\langle ad | U^t | \sigma \rangle|^2$, and the success probability in set $Y$ is $p_Y(t) = |\langle ba | U^t | \sigma \rangle|^2 + |\langle bc | U^t | \sigma \rangle|^2$, so the total success probability is $p(t) = p_X(t) + p_Y(t)$. These success probabilities are plotted in Fig.~\ref{fig:sigma_3_2} with the same parameters as Fig.~\ref{fig:s_3_2}, allowing for a direct comparison between the two initial states $\ket{s}$ and $\ket{\sigma}$. We see that $\ket{\sigma}$ behaves similarly, although it is much smoother, and the success probability in each set consistently reaches $1/2$. The jaggedness of $\ket{s}$ can be used as an advantage, however, since its success probability can have peaks greater than $1/2$. In fact, in Fig.~\ref{fig:s_3_2_3_997} and Fig.~\ref{fig:sigma_3_2_3_997} where set $X$ contains only marked vertices, the marked vertex can be found in one timestep using $\ket{s}$, whereas after one step from $\ket{\sigma}$, the success probability is still roughly $1/2$.

Now let us prove these behaviors analytically. Since we only have the eigenvectors and eigenvalues of $U$ \eqref{eq:U_bothmarked} for large $N_1$ and $N_2$ \eqref{eq:vecs_bothmarked}, we only consider this asymptotic case. Then the initial state is asymptotically
\[ \ket{\sigma} = \frac{1}{\sqrt{2}} \left( \ket{cd} + \ket{dc} \right). \]
Expressing this is a superposition of the asymptotic eigenvectors of $U$ $\eqref{eq:vecs_bothmarked}$, we get
\[ \ket{\sigma} = a\ket{\psi_1}+b\ket{\psi_2}+c\ket{\psi_3}+d\ket{\psi_4}, \]
where $a=b=-1/2$ and $c=d=1/2$. So the system approximately evolves in a smaller, 4D subspace. Applying $U^t$, taking inner products with $\bra{ab}$, $\bra{ad}$, $\bra{ba}$, and $\bra{bc}$, squaring, and adding, the success probability is
\[ p(t) = \frac{1}{2} \left[ 1 - \cos(\alpha t) \cos(\beta t) \right]. \]
Taking the derivative of this and setting it equal to zero, $p(t)$ reaches its first maximum at time $t_*$ satisfying
\[ \frac{\sqrt{k_1}}{\sqrt{N_1}}\sin\left(\frac{2\sqrt{k_1}}{\sqrt{N_1}}t_*\right)=\frac{-\sqrt{k_2}}{\sqrt{N_2}}\sin\left(\frac{2\sqrt{k_2}}{\sqrt{N_2}}t_*\right). \]
Aside from some specific cases (such as $N_1/k_1 = N_2/k_2$), there is no closed form solution to this equation. So as in the previous section with $\ket{s}$, we consider the success probability in each partite set independently.

The success probability in each partite set is
\begin{gather*}
	p_X(t) = \frac{1}{2} \sin^2 \left[ \frac{(\alpha - \beta)t}{2} \right], \\
	p_Y(t) = \frac{1}{2} \sin^2 \left[ \frac{(\alpha + \beta)t}{2} \right].
\end{gather*}
These reach their first maxima at respective times
\begin{gather*}
	t_{X*} = \frac{\pi}{\alpha-\beta} \approx \frac{\pi}{2}\sqrt{\frac{N_1}{k_1}}, \\
	t_{Y*} = \frac{\pi}{\alpha+\beta} \approx \frac{\pi}{2}\sqrt{\frac{N_2}{k_2}},
\end{gather*}
and these are the same times as for the initial state $\ket{s}$. Plugging these into $p_X(t)$ and $p_Y(t)$, each reaches a maximum of
\[ p_{X*} = p_{Y*} = \frac{1}{2}, \]
which does differ from the initial state $\ket{s}$. These results are summarized in rows four through eight of Table~\ref{table:summary}, and they mimic the results of $\ket{s}$, except now the success probability in each partite set reaches $1/2$. These results are also consistent with our numerical simulations. For example, when $N_1/k_1 = N_2/k_2$, as in Fig.~\ref{fig:sigma_3_2_600_400}, the success probability in both sets peak at the same time, resulting in a total success probability of 1.


\section{Conclusion}

We have analyzed search on the complete bipartite graph using the coined quantum walk, which is a discretization of the Dirac equation of relativistic quantum mechanics and the basis of several quantum algorithms. Although the complete bipartite graph has been considered by other forms of quantum walks, our work differs in our choice of quantum walk and initial states, which are either a uniform superposition over the vertices $\ket{s}$ or a uniform superposition over the edges $\ket{\sigma}$. Whether the marked vertices are contained in one partite set or both, the evolution with $\ket{s}$ can be jagged and greatly alternate from one timestep to the next, whereas the evolution with $\ket{\sigma}$ is much more smooth and consistent in its maximum success probability. This revealed that when using the typical initial state $\ket{s}$, care is needed to precisely obtain the runtime since being off by one timestep can negatively impact the success probability. On the other hand, this jaggedness can be exploited to improve search by identifying when the success probability is at its peaks. The overall, asymptotic results were summarized in Table~\ref{table:summary}. 

The continuous-time quantum walk, governed by Schr\"odinger's equation, searches the regular complete bipartite graph in the same runtime and with the same success probability as the complete graph, as long as $k_1 = o(N_1)$ and $k_2 = o(N_2)$. As a result, some may expect this to be true for the discrete-time coined quantum walk as well. Our work showed that this is not the case with multiple marked vertices, and so we have identified a difference between continuous-time quantum walks and discrete-time coined quantum walks.

Further research includes exploring other graphs and other initial states, and how they impact quantum walk algorithms.


\begin{acknowledgments}
	This work was partially supported by T.W.'s startup funds from Creighton University.
\end{acknowledgments}


\appendix
\section{\label{sec:appendix}Eigensystem of the Search Operator with Marked Vertices in Both Sets}

In this appendix, we find the asymptotic eigenvectors of the search operator $U$ \eqref{eq:U_bothmarked} when both partite sets contain marked vertices. For large values of $N_1$ and $N_2$, the search operator has leading terms
\[ U_0 = \begin{pmatrix}
	0 & 0 & 1 & 0 & 0 & 0 & 0 & 0 \\
	0 & 0 & 0 & 0 & 0 & 0 & -1 & 0 \\
	1 & 0 & 0 & 0 & 0 & 0 & 0 & 0 \\
	0 & 0 & 0 & 0 & -1 & 0 & 0 & 0 \\
	0 & 0 & 0 & -1 & 0 & 0 & 0 & 0 \\
	0 & 0 & 0 & 0 & 0 & 0 & 0 & 1 \\
	0 & -1 & 0 & 0 & 0 & 0 & 0 & 0 \\
	0 & 0 & 0 & 0 & 0 & 1 & 0 & 0 \\
\end{pmatrix}. \]
This has the following normalized eigenvectors and eigenvalues:
\begin{align*}
	& \ket{v_1} = \frac{1}{\sqrt{2}} \left[0,0,0,0,0,1,0,1\right],  \quad 1, \\
	& \ket{v_2} = \frac{1}{\sqrt{2}} \left[0,-1,0,0,0,0,1,0\right], \quad 1, \\
	& \ket{v_3} = \frac{1}{\sqrt{2}} \left[0,0,0,-1,1,0,0,0\right], \quad 1, \\
	& \ket{v_4} = \frac{1}{\sqrt{2}} \left[1,0,1,0,0,0,0,0\right],  \quad 1, \\
	& \ket{v_5} = \frac{1}{\sqrt{2}} \left[0,0,0,0,0,-1,0,1\right], \quad -1, \\
	& \ket{v_6} = \frac{1}{\sqrt{2}} \left[0,1,0,0,0,0,1,0\right],  \quad -1, \\
	& \ket{v_7} = \frac{1}{\sqrt{2}} \left[0,0,0,1,1,0,0,0\right],  \quad -1, \\
	& \ket{v_8} = \frac{1}{\sqrt{2}} \left[-1,0,1,0,0,0,0,0\right], \quad -1. \\
\end{align*}
The first four eigenvectors are degenerate with eigenvalue $1$, and the last four are degenerate with eigenvalue $-1$. Thus, any linear combination of the first four eigenvectors is still an eigenvector with eigenvalue $1$, and any linear combination of the last four eigenvectors is still an eigenvector with eigenvalue $-1$.

To lift the degeneracy, we include a perturbation by adding the next-order terms in $U$, resulting in
\begin{widetext}
\[ U' = \begin{pmatrix}
	0 & 0 & 1 & \frac{-2\sqrt{k_1}}{\sqrt{N_1}} & 0 & 0 & 0 & 0 \\
	0 & 0 & 0 & 0 & 0 & 0 & -1 & \frac{2\sqrt{k_1}}{\sqrt{N_1}} \\
	1 & \frac{-2\sqrt{k_2}}{\sqrt{N_2}} & 0 & 0 & 0 & 0 & 0 & 0 \\
	0 & 0 & 0 & 0 & -1 & \frac{2\sqrt{k_2}}{\sqrt{N_2}} & 0 & 0 \\
	0 & 0 & \frac{-2\sqrt{k_1}}{\sqrt{N_1}} & -1 & 0 & 0 & 0 & 0 \\
	0 & 0 & 0 & 0 & 0 & 0 & \frac{2\sqrt{k_1}}{\sqrt{N_1}} & 1 \\
	\frac{-2\sqrt{k_2}}{\sqrt{N_2}} & -1 & 0 & 0 & 0 & 0 & 0 & 0 \\
	0 & 0 & 0 & 0 & \frac{2\sqrt{k_2}}{\sqrt{N_2}} & 1 & 0 & 0 \\
\end{pmatrix}. \]
\end{widetext}
We look for eigenvectors of this that are linear combinations $\alpha_1 \ket{v_1} + \alpha_2 \ket{v_2} + \alpha_3 \ket{v_3} + \alpha_4 \ket{v_4}$, so they satisfy
\[ \begin{pmatrix}
	U'_{11} & U'_{12} & U'_{13} & U'_{14} \\
	U'_{21} & U'_{22} & U'_{23} & U'_{24} \\
	U'_{31} & U'_{32} & U'_{33} & U'_{34} \\
	U'_{41} & U'_{42} & U'_{43} & U'_{44} \\
\end{pmatrix} \begin{pmatrix} \alpha_1 \\ \alpha_2 \\ \alpha_3 \\ \alpha_4 \end{pmatrix} = \lambda \begin{pmatrix} \alpha_1 \\ \alpha_2 \\ \alpha_3 \\ \alpha_4 \end{pmatrix}, \]
where $U'_{ij} = \langle v_i | U' | v_j \rangle$. Evaluating the matrix elements,
\[ \begin{pmatrix}
	1 & \sqrt{\frac{k_1}{N_1}} & \sqrt{\frac{k_2}{N_2}} & 0 \\
	-\sqrt{\frac{k_1}{N_1}} & 1 & 0 & -\sqrt{\frac{k_2}{N_2}} \\
	-\sqrt{\frac{k_2}{N_2}} & 0 & 1 & -\sqrt{\frac{k_1}{N_1}} \\
	0 & \sqrt{\frac{k_2}{N_2}} & \sqrt{\frac{k_1}{N_1}} & 1 \\
\end{pmatrix} \begin{pmatrix} \alpha_1 \\ \alpha_2 \\ \alpha_3 \\ \alpha_4 \end{pmatrix} = \lambda \begin{pmatrix} \alpha_1 \\ \alpha_2 \\ \alpha_3 \\ \alpha_4 \end{pmatrix}. \]
This has solutions
\begin{align*}
	& \ket{\psi_1} = \frac{1}{2} \left( -\ket{v_1} - i \ket{v_2} + i \ket{v_3} + \ket{v_4} \right), \quad \lambda_1=e^{-i\alpha}, \\
	& \ket{\psi_2} = \frac{1}{2} \left( -\ket{v_1} + i \ket{v_2} - i \ket{v_3} + \ket{v_4} \right), \quad \lambda_2=e^{i\alpha}, \\
	& \ket{\psi_3} = \frac{1}{2} \left(  \ket{v_1} - i \ket{v_2} - i \ket{v_3} + \ket{v_4} \right), \quad \lambda_3=e^{-i\beta}, \\
	& \ket{\psi_4} = \frac{1}{2} \left(  \ket{v_1} + i \ket{v_2} + i \ket{v_3} + \ket{v_4} \right), \quad \lambda_4=e^{i\beta},
\end{align*}
where
\[ \sin\alpha=\sqrt{\frac{k_2}{N_2}}+\sqrt{\frac{k_1}{N_1}}, \quad \sin\beta=\sqrt{\frac{k_2}{N_2}}-\sqrt{\frac{k_1}{N_1}}. \]
Substituting in for $\ket{v_1}, \dots, \ket{v_4}$, we get the first four eigenvectors in the main text \eqref{eq:vecs_bothmarked}.

Similarly, we look for eigenvectors of $U'$ that are linear combinations $\alpha_5 \ket{v_5} + \alpha_6 \ket{v_6} + \alpha_7 \ket{v_7} + \alpha_8 \ket{v_8}$, so they satisfy
\[ \begin{pmatrix}
	U'_{55} & U'_{56} & U'_{57} & U'_{58} \\
	U'_{65} & U'_{66} & U'_{67} & U'_{68} \\
	U'_{75} & U'_{76} & U'_{77} & U'_{78} \\
	U'_{85} & U'_{86} & U'_{87} & U'_{88} \\
\end{pmatrix} \begin{pmatrix} \alpha_5 \\ \alpha_6 \\ \alpha_7 \\ \alpha_8 \end{pmatrix} = \lambda \begin{pmatrix} \alpha_5 \\ \alpha_6 \\ \alpha_7 \\ \alpha_8 \end{pmatrix}. \]
Evaluating the matrix elements,
\[ \begin{pmatrix}
	-1 & -\sqrt{\frac{k_1}{N_1}} & \sqrt{\frac{k_2}{N_2}} & 0 \\
	\sqrt{\frac{k_1}{N_1}} & -1 & 0 & \sqrt{\frac{k_2}{N_2}} \\
	-\sqrt{\frac{k_2}{N_2}} & 0 & -1 & -\sqrt{\frac{k_1}{N_1}} \\
	0 & -\sqrt{\frac{k_2}{N_2}} & \sqrt{\frac{k_1}{N_1}} & -1 \\
\end{pmatrix} \begin{pmatrix} \alpha_5 \\ \alpha_6 \\ \alpha_7 \\ \alpha_8 \end{pmatrix} = \lambda \begin{pmatrix} \alpha_5 \\ \alpha_6 \\ \alpha_7 \\ \alpha_8 \end{pmatrix}. \]
This has solutions
\begin{align*}
	& \ket{\psi_5} = \frac{1}{2} \left( -\ket{v_5} + i \ket{v_6} + i \ket{v_7} + \ket{v_8} \right), \quad \lambda_5=-e^{i\alpha}, \\
	& \ket{\psi_6} = \frac{1}{2} \left( -\ket{v_5} - i \ket{v_6} - i \ket{v_7} + \ket{v_8} \right), \quad \lambda_6=-e^{-i\alpha}, \\
	& \ket{\psi_7} = \frac{1}{2} \left(  \ket{v_5} + i \ket{v_6} - i \ket{v_7} + \ket{v_8} \right), \quad \lambda_7=-e^{i\beta}, \\
	& \ket{\psi_8} = \frac{1}{2} \left(  \ket{v_5} - i \ket{v_6} + i \ket{v_7} + \ket{v_8} \right), \quad \lambda_8=-e^{-i\beta}.
\end{align*}
Substituting in for $\ket{v_5}, \dots, \ket{v_8}$, we get the last four eigenvectors in the main text \eqref{eq:vecs_bothmarked}.
\vspace{1in}


\bibliography{refs}

\end{document}